\documentclass{article}
\usepackage[T1]{fontenc}
\usepackage{wrapfig}
\usepackage[portuges,english]{babel}
\usepackage{amssymb,latexsym,amsmath,color,mathrsfs,ifsym,graphics,stmaryrd} %stmaryrd for \llbracket and \rrbracket
\usepackage[colorlinks,linkcolor=blue,urlcolor=blue,citecolor=black,
plainpages=false,pdfpagelabels,breaklinks]{hyperref}

\title{Revisiting the Applicability of\\Metaphysical Identity in Quantum Mechanics}

\author{{\sc N. C. A. da Costa}$^{1}$\ {\sc and  C. de Ronde}\ $^{2}$}
\date{}

\usepackage[margin=3.5cm]{geometry}

\begin{document}

\bibliographystyle{plain}
\maketitle

\begin{center}
\begin{small}
1. Federal University of Santa Catarina - Brazil \\
2. CONICET, University of Buenos Aires  - Argentina \\
Center Leo Apostel and Foundations of  the Exact Sciences\\
Brussels Free University - Belgium \\
\end{small}
\end{center}

\begin{abstract}
\noindent We discuss the hypothesis that the debate about the interpretation of the orthodox formalism of quantum mechanics (QM) might have been misguided right from the start by a biased metaphysical interpretation of the formalism and its inner mathematical relations. In particular, we  focus on the orthodox interpretation of the {\it congruence relation}, `=', which relates {\it equivalent classes} of different mathematical representations of a vector in Hilbert space, in terms of {\it metaphysical identity}. We will argue that this seemingly ``common sense'' interpretation, at the semantic level, has severe difficulties when considering the syntactic level of the theory.   

\end{abstract}

\noindent\textbf{Keywords}: metaphysical identity, congruence relation, interpretation of quantum mechanics.

\bigskip 

%--------------------------------------------------------------
\renewenvironment{enumerate}{\begin{list}{}{\rm \labelwidth 0mm
\leftmargin 0mm}} {\end{list}}

\newcommand{\ita}{\textit}
\newcommand{\mcal}{\mathcal}
\newcommand{\mfrak}{\mathfrak}
\newcommand{\mbb}{\mathbb}
\newcommand{\mrm}{\mathrm}
\newcommand{\msf}{\mathsf}
\newcommand{\mscr}{\mathscr}
\newcommand{\lra}{\leftrightarrow}
\renewenvironment{enumerate}{\begin{list}{}{\rm \labelwidth 0mm
\leftmargin 5mm}} {\end{list}}

\newtheorem{dfn}{\sc{Definition}}[section]
\newtheorem{thm}{\sc{Theorem}}[section]
\newtheorem{lem}{\sc{Lemma}}[section]
\newtheorem{cor}[thm]{\sc{Corollary}}
\newcommand{\Proof}{\textit{Proof:} \,}
\newcommand{\cqd}{{\rule{.70ex}{2ex}} \medskip}

\section*{Introduction}

In this paper we question whether the so called ``minimal interpretation'' of the formalism of quantum mechanics (QM) might have been misguided by a biased metaphysical interpretation of the mathematical formalism and its inner relations. This goes clearly against the idea according to which it is possible to ``read literally'' an interpretation of QM from the formalism alone, or in other words, that the mathematical formalism of QM has a ``transparent'' metaphysical (or conceptual) reading. In particular, we will analyze the general interpretation given to the equality sign `$=$' in the orthodox quantum formalism which relates different mathematical representations of a vector in Hilbert space. 

We will argue that a mathematical formalism alone is not constrained by any sort of metaphysical commitment. In physical theories, the metaphysical understanding must be added to the formalism in such a way that specific physical concepts provide a coherent representation of what is going on according to the theory. The uncritical application of inadequate concepts to an abstract formalism might produce not only a misunderstanding of the mathematical structure but also different pseudoproblems. This might be the case within the foundational debates about QM where the orthodox formalism has been understood in terms of metaphysical identity. Contrary to this widespread (metaphysical) interpretation, we will provide arguments which show that the {\it equivalent classes} of a vector in Hilbert space (related through the {\it congruence relation}, `$=$') cannot, in general, be regarded as making reference to a {\it metaphysical identity}. Even though we remark that we are not discussing here the well known problem of identical or indistinguishable particles in QM (see \cite{ArenhartKrause12, daCostadeRonde14, FrenchKrause06}), our analysis will have deep consequences for that debate which we attempt to discuss in a future work. 

The paper is organized as follows. In section 1 we introduce and discuss the axiomatization and meaning of formal theories as related to metaphysical schemes, in particular we will consider the notion of identity. In section 2 we discuss the notion of identification within mathematical structures and relate it to metaphysical identity. Section 3 analyses the relation between logical and metaphysical principles. In section 4 we distinguish the syntactic and semantic levels of a theory making special emphasis on the fact that while the former level is symbolic and abstract, it is the latter level which introduces a language that implies a specific metaphysical or conceptual interpretation of such symbols. Section 5, discusses the application of metaphysical identity to QM and the problems it gives rise to when considering the contextual relation between multiple mathematical representations of a vector. 

\section{Axiomatization, Meaning and Identity} 

In certain sense, the meaning of a formalized or only partially formalized theory $T$ is implicitly given by the axiomatic systematization of $T$. The primitive concepts and primitive propositions of $T$, plus its underlying logic, fix the possible models of $T$ and the possible meanings of its primitive concepts and of its primitive propositions (axioms). The remaining concepts of $T$ have to be explicitly introduced by explicit (nominal) definitions. In fact, the axiomatization of $T$ constitutes an implicit definition of the possible systems of objects of which the theory talks about. $T$ has, in principle, various (infinite) different interpretations, that is, logically different models. In consequence, the primitive concepts and propositions of $T$ may acquire various special, primitive, meanings.\footnote{Notice that there two senses of ``meaning'': meaning provided by the axiomatization, that is, by the syntactical structure of the theory, and meaning as attributed by an interpretation. The two are obviously not equal.} The same also occurs when $T$ is extended to several stronger theories. Therefore if the meaning of a theory is determined by its axiomatization, its deductive structure, then in the stronger theory the meanings of the corresponding concepts need to differ from those of the initial theory. Diverse axiomatizations entail unlike kinds of meaning. 

Commonly, the above view of the meaning of theories is not extended to logic itself. While the meanings of the specific notions of theories are, so to say, relative, the logical notions are absolute, and remain strictly the same in all interpretations (of the primitive concepts and the axioms). Our main thesis is that the central notions of a logical system ${\cal L}$, for instance the classical predicate calculus of first order with equality, has diverse meanings, which depend on the theory whose logic is ${\cal L}$. That is, the meaning has a global signification, being in some sense relative. For example, the meaning of the quantifiers in a standard mathematical theory and their meaning in QM, are globally different. The standard conception of the logical notions, as fixed and immutable, can only be uphold by an appeal to a metaphysical stance.\footnote{In connection with non-classical logics, the global aspect of meaning seems clearer.} 

All the above considerations apply in particular to the case of the concept of mathematical identity (or equality). However the situation of this kind of notion presents an extra complication, that we now discuss. To begin with, {\it mathematical identity} is an equivalence relation, and if a theory $T$ contains the equality symbol `$=$', it may refer to some equivalence relation and not to a {\it metaphysical identity}. However, if $T$ possesses an equivalence relation $\equiv$, it has also an interpretation of $\equiv$ as a possible metaphysical identity. Concequently, if $T$ does contain this second kind of identity, then it has an interpretation according to which the metaphysical identity is, in fact, only an equivalence relation. In general, we don't have at our disposition a process to distinguish these situations. Formally, we are unable to characterize the concept of identity; analogously, if we try to employ semantic methods, say standard semantics, we are supposed to apply set-theoretic methods that are formulated inside a set theory, and the problem reappears.

In this paper we attempt to distinguishing {\it metaphysical identity} from {\it equivalence relations}. We will show that if we are not rigorous in this respect, we surely will commit significant equivocations and mistakes. In fact, the orthodox interpretation of QM might have produced this kind of error by uncritically presupposing an interpretation of mathematical {\it equivalence classes} in terms of a {\it metaphysical identity}.  

Identity is a `strong' equivalence relation. So, it is supposed to be reflexive, symmetric, transitive and to satisfy the replacement scheme:
\begin{equation}
x = y \rightarrow (F(x)  \leftrightarrow F(y))
\end{equation}

\noindent where the notations have the usual meanings. We are restricting the analysis to classical first-order logic with identity, and we will denote by $L$ such a logic and by ${\cal L}$ its language. All instances of the scheme or principle of replacement belong to ${\cal L}$ and this scheme is, therefore, language dependent; consequently, if we change the language, the principle above also changes. In other words, the global meaning of identity is dependent of the language employed. 

Let  ${\cal I}$ be an interpretation of the language  ${\cal L}$ of a (first-order) theory $T$, whose logic is $L$. $T$ divides the domain of  ${\cal I}$ in equivalence classes as follows: two members of the domain, $x$ and $y$, belong to the same equivalence class if and only if 
\begin{equation}
F(x)  \leftrightarrow F(y)
\end{equation}

\noindent is true according to  ${\cal I}$ for all formulas $F$ of the language  ${\cal L}$, under clear qualifications. It is easy, then, to show that the equivalence classes constitute the basic domain of an interpretation  ${\cal I'}$ which is a model of $T$, when appropriate definitions are employed. Procedures like this demonstrate that if in an interpretation of a theory the symbol of identity is interpreted as the metaphysical identity, then there is another interpretation in which the same symbol denotes a strong equivalence relation, and conversely. 

We make here only some references concerning higher-order predicate logic: 

\begin{enumerate}
\item[i.] The notion of mathematical identity is delimitated by a theory of types. Otherwise, we can reproduce the Russell paradoxes for predicates and relations. 

\item[ii.] The semantics of higher-order logics normally presuppose set theory, which is a first-order theory, and we meet again the relative character of identity.

\item[iii.] The logical semantics of a given language is developed in a meta-language which includes identity and in its rigorous formulation doesn't fix the meaning of identity. 
\end{enumerate}

\noindent We arrive at the conclusion that in first-order or in higher-order logic it is not possible to characterize, univocally, the concept of {\it absolute identity}. In order to do that, we must appeal to metaphysical principles. In fact, the analysis and discussions of principles like identity, the indistinguishability of identicals, and the identity of indistinguishables, involve us into deep metaphysical problems, with repercussions throughout logic and physics. In this paper we will apply this analysis to the specific case of QM, where the distinction between the formal and metaphysical (or interpretational) levels of discourse have been presupposed in the orthodox literature right from the start. 

The preceding discussion could be extended to cover most systems of non classical logic, for instance, intuitionistic logic and paraconsistent logic. However this will be left to a future work.

\section{Formal Identification}

Identification is an operation that belongs to the theory of mathematical structures. We present here, for the sake of completeness, a rough outline of the concept of (mathematical) structure (see \cite{Bourbaki68} and \cite{daCostaBueno15}). A scale of sets on a finite number of initial sets is a finite sequence of sets, such that each of them is an initial set or is obtained from preceding sets of the sequence by forming power sets or cartesian products, according to a scheme of formation. Let $M$ be a scale of sets based on, say, sets $A$, $B$ and $C$; let us suppose that we have formulated a list of propositions, called axioms, expressing set theoretic properties of a generic element of $M$, and let us, also, denote by $T$ the intersection of the subsets of $M$ defined by such properties. Any element of $T$ defines a structure of species $T$ on $A$, $B$ and $C$. Therefore a structure is determined by the scheme of $M$, from the base sets $A$, $B$ and $C$, and the axioms of the structure. Groups, topological spaces, fields, Hilbert spaces, von Neumann algebras, real numbers, complex numbers, etc., are all of them mathematical structures. Further, it is not difficult to define a product of structures, isomorphism of structures, etc. 

Bourbaki introduces the concept of {\it identification} as follows, considering only structures with one base set (but it is easy to extend the definition to the case of various sets): 

\begin{quotation}
\noindent {\small ``When there exists an isomorphism $f$ of a set $E$, endowed with a structure $G$, onto a set $E'$, endowed with a structure $G'$, it is often convenient to {\it identify} $E$ with $E'$, i.e., to give {\it the same name} to an element of a set $M$ in the scale based on $E$ and to the element which is its image under the appropriate extension of $f$ to the set $M$.'' \cite[p. 385]{Bourbaki68}}\end{quotation} 

All these ideas may be found in the {\it Summary of Results} of Bourbaki \cite[pp. 347-385]{Bourbaki68}. In the bulk of his book, Bourbaki presents a different concept of structure, more general and more sophisticated. However, the notion really employed in common mathematics is the one of the {\it Summary}. There is yet another concept of mathematical structure introduced by Ch. Ehremann, the `functorial' concept of structure (see \cite{Corry96}). A careful study of identification in topological spaces is contained in \cite[pp. 42-43]{Hu65}.

Classical mathematics is based on set theory and its central objects are mathematical structures. Taking into account the above considerations, mathematical identity may be interpreted as a strong equivalence relation or congruence, and not only as a metaphysical identity. In particular, if two structures are {\it mathematically identical}, this does not mean that they are also {\it metaphysically identical}. The objects of classical mathematics are, fundamentally, structures; from the abstract point of view, the objects of classical mechanics and of quantum mechanics  are structures. So, when identity appears in these two different physical theories, it is not, necessarily, {\it absolute identity}, which implies a metaphysical interpretation, but simply a {\it congruence relation}. In this way, from the abstract point of view, equality or identity, in the theory of Hilbert spaces can be seen as a congruence. Therefore, the equality expressing the decomposition of a vector in some base, may also constitute a congruence.

All those remarks are reinforced by the true meaning of the identification in connection with structures. Identification, so common in pure mathematics, usually refers to a question of designation, a convention concerning the employment of names. For instance, the real numbers, conceived either as sequences of rationals, or as cuts {\it \`a la} Dedekind in the field of rational numbers, are identified. Thus they are, so to say, ``the same thing''. So, {\it identifcation} in mathematics has nothing to do with the notion of {\it metaphysical identity}. 

Now that we have discussed identity within formal theories we will turn our attention in the following section to the more general relation between logical and metaphysical principles.

\section{Logical Identity and Metaphysical Identity}

Logic and physics have been intimately related since their origin. It was Aristotle who created classical logic and used it in order to develop his own physical and metaphysical scheme, providing a solution to the problem of movement and knowledge set down ---according to both Plato and Aristotle--- by the Heraclitean and Eleatic schools of thought. Movement was then accounted by Aristotle in terms of his hylomorphic scheme, as the path from a potential (undetermined, contradictory and non-identical) realm to an actual (determined, non-contradictory and identical) realm of existence. The notion of entity was then characterized by three main logical and ontological principles: the Principle of Existence (PE) which allowed Aristotle to claim existence about that which is predicated, the Principle of Non-Contradiction (PNC) which permitted him to argue that the existent possesses non-contradictory properties, and the Principle of Identity (PI) which allowed him to claim that the predicated non-contradictory existent is ``the same'', or remains ``identical to itself'', through time. As Verelst and Coecke make the point:

\begin{quotation}
\noindent {\small ``The three fundamental principles of classical (Aristotelian) logic: the existence of objects of knowledge, the principle of contradiction and the principle of identity, all correspond to a fundamental aspect of his ontology. This is exemplified in the three possible usages of the verb {\it to be}': existential, predicative, and identical. The Aristotelian syllogism always starts with the affirmation of existence: something is. The principle of contradiction then concerns the way one can speak (predicate) validly about this existing object, i.e. about the true and falsehood of its having properties, not about its being in existence. {\it The principle of identity states that the entity is identical to itself at any moment (a=a), thus granting the stability necessary to name (identify) it.}'' \cite[p. 172]{VerelstCoecke} (emphasis added)}
\end{quotation}

\noindent PI made possible ---at the metaphysical level--- the constitution of the notion of entity as ``something'' which is metaphysically identical to itself. It is this architectonic which allows all particular representations of an entity to be related to itself, as a unity, as a sameness. However, regardless of its common origin and deep interrelation, it should be clear that even though logic and metaphysics might be interrelated, logical schemes are not necessarily nor uniquely related to particular metaphysical schemes. This is not a ``self evident'' relation. 

For many centuries it was dogmatically assumed that classical logic was the ``true logic'' of our world, the only guide of rational thought, the formal apparatus for correct reasoning. However, the revolution in the foundations of mathematics and logic that took place at the beginning of the 20th Century produced a whole new group of non-classical logics. Paraconsistent logics, non-reflexive logics, fuzzy logics, etc., where all developed breaking the exclusive ruling of the Aristotelian principles (see \cite{daCostaBueno09, daCostaKrauseBueno07, Zadeh75}). Thus it became clear that the fundament of ``correct'' reasoning was neither firm nor ``self evident''. This proliferation of logics made also possible to understand more clearly the complex relation between formal languages and physical theories. In fact, the revolution that took place in logic has a clear counterpart in the foundations of physics where also new theories were created with different metaphysical standpoints to those of classical physics. While Relativity theory deconstructed classical Newtonian absolute space and time and produced a new Riemannian entangled space-time, QM presented serious difficulties to interpret the formalism in terms of ``classical reality'' ---i.e., in terms of PE, PNC and PI.

We now turn our attention to the important distinction, within theories, between syntax and semantics. This distinction will allow us to make explicit, in the following sections, the problem we attempt to expose regarding the applicability of metaphysical identity within the orthodox interpretation of QM.

\section{Syntactic and Semantic Levels}

Any theory $T$ is built in a language $L$. $L$ has two basic dimensions: The syntactic and the semantic. The former concerns the symbolic, syntactic part of $L$; it constitutes a kind of geometry of symbols. The latter deals with the interpretation of $L$ and, indirectly, of $T$. The semantics of $T$ reflects, then, a particular interpretation of $T$ (and of $L$) including the denotation of its terms and the truth of its sentences. Consequently, the symbol for identity, that $T$ is supposed to possess, presents also two levels: the syntactical and the semantical levels. The usual syntactic properties of identity are reflexivity, symmetry, transitivity and substitution, as mentioned above. However, these fundamental properties don't characterize identity completely, since the theory $T$, that contains an identity symbol, has semantic interpretations in which this symbol can be interpreted in different ways. 

There exists, so to say, a kind of pluralism concerning the interpretation of identity and, as we already noted, this pluralism encompasses all logical notions which, leaving aside a clearly metaphysical stance, do not have a unique, possible interpretation (and, therefore, meaning). There exists no (strong) argument to rule out the metaphysical interpretation of identity in most cases, in particular in logic and mathematics. However, the usual properties of the identity symbol are not enough to guarantee that its interpretation should be {\it necessarily} provided in terms of metaphysical identity. In effect, a theory like $T$, as mentioned above, based on first-order logic with identity, if consistent, has models in which identity may be interpreted as a strict {\it congruence}; i.e., it isn't the relation of metaphysical identity. The same remains true if the underlying logic of $T$ is higher order logic or set theory. 

The choice of a specific language implies also the choice of a particular semantic and metaphysical scheme. Language, semantics and metaphysics come all together. Every notion of the language must be rigorously defined for the lack of such a precise definition of the meaning of notions (e.g. `individual', `object', `state', etc.) would preclude the very possibility of analysis of such notions. These definitions are not in the syntactical level, they are in the semantical one. As we have discussed above, the relation between a mathematical structure, in the syntactic level, and a metaphysical conceptual scheme, in the semantic level, is not a ``self evident'' relation or something which can be ``literally read'' from the syntactic level. 

Each theory $T$ requires the development and creation of specific interrelations between the syntactical and semantical levels. One could argue, following an instrumentalist perspective, that such a relation is completely unnecessary in the case of physical theories. In fact, there is no consensus within the specialized literature regarding the question over the necessity or not to interpret mathematical formalisms like that of QM (e.g. \cite{VassalloEsfeld15, Marchildon04}). For example, Fuchs and Peres \cite[p. 70]{FuchsPeres00} have argued, in a paper entitled {\it Quantum Theory Needs No `Interpretation'}, that ``[...] quantum theory does not describe physical reality. What it does is provide an algorithm for computing probabilities for the macroscopic events (``detector clicks'') that are the consequences of experimental interventions. This strict definition of the scope of quantum theory is the only interpretation ever needed, whether by experimenters or theorists.'' This instrumentalist perspective is satisfied with having an ``algorithmic recipe'' that allows the physicist to calculate measurement outcomes from the formalism. Here, the semantic level seems to require a minimum of structure since there is no need to supplement the theory with an interpretation that would conceptually explain its relation to physical reality. However, in this case the relation between the syntax of a theory and experience becomes highly problematic. In fact, abstract mathematical symbols provide no guide of how to interpret them in terms of physical notions. For example, the theory of calculus does not contain the physical notions of Newtonian mechanics. One cannot find ``hidden'' between the abstract symbols of calculus the physical concepts of `space', `time', `inertia', `mass'', etc. 

On the very contrary, we view it that one of the basic tasks of the physicists and philosophers of physics is to produce theories which provide physical representations which allow us to grasp and understand the world around us. In order to provide such formal-conceptual representation physicists must necessarily complement mathematical formalisms, in the syntactic level, with networks of physical concepts, in the semantic level. This is the reason why, unless a form of Platonism is taken up whereby mathematics is ontology, it is simply not enough to claim that ``according to QM the structure of the world is like Hilbert space'' or that ``the quantum state $\Psi$ describes physical reality''. That is just mixing the semantic and syntactical levels of the theory. That is not doing the work of providing a conceptual representation in the sense discussed above. And that is the reason why all physical theories require {\it necessarily} a semantic level. As we argued above,  this semantic level is defined not only linguistically and conceptually but also metaphysically.  

In the previous sections we discussed how the notion of entity in classical physics is defined and constrained by Aristotelian logic and metaphysics through its principles: PE, PNC and PI. However, as we all know, QM faces severe difficulties to relate its formalism to the constraints imposed by these classical principles. In the following section we attempt to show how metaphysical identity has been uncritically applied to vectors in Hilbert space erasing the richness of the mathematical structure of the theory.

\section{Applying Metaphysical Identity to QM}

As we argued above, an entity\footnote{The notions of `entity', `system', `object' play the role of synonyms. These notions make reference to a specific way of understanding physical reality.} in classical physics is always ---implicitly or explicitly--- metaphysically defined in terms of PE, PNC and PI.  Without such a rigorous (formal and metaphysical) definition, the notion of `physical system' losses not only its physical content, but also its applicability. Indeed, without PE it is not possible to make reference to the existence of a system; without PNC it becomes impossible to describe such system in terms of `properties'; and without PI it would not be possible to follow the system through time ---turning `evolution' itself into a meaningless physical notion. Even though it is difficult to find such an explicit definition within part of a literature which attempts to escape metaphysical debates, there is always space for implicit definitions. For example, Dieks and Lubberdink make use of these principles when they discuss the meaning of the indistinguishability of ``classical particles'': 

\begin{quotation}
\noindent {\small``In classical physics, particles are the example {\it par excellence} of distinguishable individuals. No two classical particles can {\it be} [assuming PE] in exactly the same physical state: in Newtonian spacetime different particles will at least occupy different spatial positions at any moment, because of their impenetrability. They will therefore obey Leibniz's Principle of the Identity of Indiscernibles, which says that different individuals [assuming PNC] cannot share all their physical properties. Moreover, classical particles possess {\it genidentity}, i.e. {\it identity} [of non-contradictory properties, PNC] {\it over time} [assuming PI].'' \cite[pp. 2-3]{DieksLubberdink11}}
\end{quotation}

\noindent The orthodox interpretation of QM assumes that a vector in Hilbert space ---in analogous fashion to the interpretation of a point in phase space in classical mechanics--- represents the state of a `quantum system'. Of course this interpretation must be accompanied by a {\it coherency requirement} regarding the sound applicability of PE, PNC and PI within the quantum formalism. In case we want to talk about a system in a sensible manner within the theory, the quantum system should exist (PE), possess non-contradictory properties (PNC) and remain identical to itself through time (PI). 

In the case of QM the application of identity to the formalism has produced the idea that a quantum state is the same as a superposition. The notions of `pure state', `quantum particle', `quantum system' and `quantum superposition' have been scrambled confusing not only different mathematical representations of a vector but also different levels of mathematical representation ---levels which have remained unnoticed by philosophers of QM. More importantly, this state of turmoil has made it impossible to discuss the meaning of quantum superpositions themselves. In fact, there is no specific notion in the literature to pin down each one of these distinct contextual elements of the theory. Let us explain this in more detail. 

It is common to find within any paper which discusses about QM, in order to present a superposition state the following equation: 
\begin{equation}
| \psi\rangle = N \ (|a_1\rangle +  \ |a_2\rangle) 
\end{equation}

\noindent Both sides of the equation are interpreted as describing {\it the same} quantum system or individual. Each one of the kets $(| \psi\rangle, |a_1\rangle, |a_2\rangle)$ is related to projection operators ($| \psi\rangle\langle \psi |$, $| a_1\rangle\langle a_1 |$, $| a_2 \rangle\langle a_2 |$) which in turn are interpreted as properties which pertain to the quantum system. The ---implicit or explicit--- idea behind this interpretational choice is that a quantum system $| \psi\rangle$ possess in some way all the properties that can be found within its multiple representations in terms of linear combinations. Already here there is a confusion for both terms of Equation (3) are mathematically equivalent. They are in the same level of mathematical representation ---both are representations of vector in a particular basis. So why do we find this equation in most papers which discuss about quantum superpositions or quantum systems? Why does everyone write $| \psi\rangle$ in the left hand side of the equation and  $N \ (|a_1\rangle +  \ |a_2\rangle)$ ---or any other superposed state--- in the right hand side of the equation? Is a superposition of one term more important than one of many terms? Is the left hand side term containing in some way the term in the right hand side? In fact, through a change of basis we could also write the following equivalence: 
\begin{equation}
N \ (|a_1\rangle +  \ |a_2\rangle) = M \ (|b_1\rangle +  \ |b_2\rangle) 
\end{equation}

\noindent These obvious remarks expose a deep metaphysical presupposition assumed within the orthodox literature which is simply not contained in the mathematical formalism of the theory. We conjecture that this ``reading'' of the formalism goes back to a deep metaphysical idea of our Western culture, discussed by the ancient Greeks, namely, the relation between {\it the one and the many}. The idea that {\it the one} contains {\it the multiple}. An object contains its properties. That might be the metaphysical scheme behind the choice of writing equation Eq. (3) over of Eq. (4). However, regardless of such metaphysical orthodox presuppositions, this is not what Eq. (3) is talking about.

$| \psi\rangle$ does not represent an object nor a system according to the orthodox formalism. In the orthodox interpretation $| \psi\rangle$ is interpreted as a ``state'' of the quantum system which relates to the projection operator $| \psi\rangle \langle \psi |$, which in turn is interpreted as a property. $N \ (|a_1\rangle +  \ |a_2\rangle)$ is just another ``state''. The only specific mathematical characteristic of $| \psi\rangle$ is that it is a superposition of only one term. But apart from the fact that our culture gives a deep importance to the number 1, there is no reason why to consider $| \psi\rangle$ as ``more important'' or ``special'' than any other mathematical representation of the same vector. Since both $| \psi\rangle$ and $N \ (|a_1\rangle +  \ |a_2\rangle)$ are equivalent mathematical representations of the same vector, there is no formal reason that would allow us to justify why we must write Eq. (3) instead of writing directly the superposition $N \ (|a_1\rangle +  \ |a_2\rangle)$. 

We begin to understand how a particular interpretation has been imposed without respecting the symmetries and features present within the quantum syntax. The {\it congruence relation} `=' interpreted in terms of {\it metaphysical identity} is implicitly  considered as making reference to `the same quantum system'. Through the notion of `quantum system', and also that of `pure state', {\it metaphysical identity} is implicitly applied to the mathematical formalism. This presupposition allows us to read the two terms of Eq. (3) as making reference to {\it the same} quantum individual. In order to reinforce this idea it has become a habit in the literature to always write a superposition of many terms in relation to a superposition of one term. In fact, this is the way the notion of superposition is taught to physicists in Universities all around the world. As we shall see, this metaphysical choice has made almost impossible the debate about the meaning of quantum superpositions beyond classical metaphysics. Even the most obvious fact about the paraconsistent character of quantum superpositions has been completely overseen in the literature and remarked only recently by the same authors of this paper \cite{daCostadeRonde13}.  

In order to make explicit the distinction between the two different quantum superpositions in Equations (3) and (4), in \cite{deRonde16b} one of us has proposed to consider these different mathematical representations through an explicit contextual definition of the notion of quantum superposition which considers right from the start  their obvious basis-dependent character. This is only a first step in order to escape the orthodox ``common sense'' semantics imposed by classical metaphysics. Let us recall the contextual definition of quantum superpositions: 

\begin{dfn}
{\sc Quantum Superposition:} Given a quantum state, $\Psi$, each i-basis defines a particular mathematical representation of $\Psi$, $\sum c_i \ | \alpha_i \rangle$, which we call a quantum superposition. The notion of quantum superposition is contextual for it is always defined in terms of a particular context (or basis).
\end{dfn}

\noindent This definition has not been made explict in the orthodox literature due to to the application of metaphysical identity within the interpretation of a vector in Hilbert space. The abstract vector $\Psi$  is equated with a particular mathematical representation given by a superposition of one element $| \psi\rangle$. This, we argue, is a mistake.

The application of metaphysical identity hides two very important distinctions when attempting to interpret the formalism. Firstly, the distinction between different quantum superpositions; and secondly, the distinction ---to which we now turn--- between the {\it one} abstract vector, $\Psi$, and its {\it multiple} basis dependent mathematical representations, $\sum c_i \ | \alpha_i \rangle$. 

From a formal perspective, all bases have exactly the same importance. The ket $| \psi\rangle$ is just a particular mathematical representation given by the choice of a specific basis of the abstract vector $\Psi$. From a formal perspective, as we remarked above, there is no reason to consider this basis as ``more important'' or ``special'' than any other basis. There is in fact an infinite number of $i$-representations of this vector which depend on the choice of the particular $i$-basis. In this case, the one abstract vector $\Psi$ contains its multiple representations. But $\Psi$ is not $| \psi\rangle$. In order to make explicit the two different levels, which have been mixed in the orthodox literature, we can write the following equation: 
\begin{equation}
\Psi = N \ (|a_1\rangle +  \ |a_2\rangle) = M (|b_1\rangle +  \ |b_2\rangle) = | \psi\rangle = ... = K (|r_1\rangle +  \ |r_2\rangle)
\end{equation}

The important point we want to remark is that the semantic interpretation of Equation (5) is not at all ``obvious'' nor ``self evident''. One cannot ``read literally'' the existence of a `system' or of `properties' from this equation. The orthodox interpretation which assumes that all these mathematical representations are making reference to {\it the same} quantum system is certainly not the only possible interpretation that one could produce. As we remarked above, the idea that all the terms of Equation (5) are considered to be {\it the same} becomes evident when we recall there does not exist a specific notion, within the orthodox interpretation, that would allow us to distinguish between the different terms. This is done irrespectively of the contextual character of QM which points exactly in the opposite direction, namely, to the fact there exists an important specificity regarding each particular expression.  

Irrespectively of quantum contextuality, Equation (5) has been interpreted through the choice of a language ---following classical metaphysics--- containing noncontextual notions like `system' and `property'. What might seem as ``harmless'' or even ``self evident'' presupposition is in fact a strong metaphysical choice which applies {\it metaphysical identity} where, in fact, there is only a {\it congruence relation}. As we have shown, the danger of the uncritical choice of a language (a semantic) ---which imposes definite metaphysical commitments--- comes from the limit it imposes on the very possibilities of analysis. In this respect, we might also recall the old positivist lesson according to which the use of inadequate concepts within a language can only lead to the creation of pseudoproblems. 

In orthodox QM {\it metaphysical identity} is applied implicitly. This makes it very difficult to debate about the adequacy of its applicability. Fortunately for us, there is an interpretation which not only makes an explicit use of this principle but also derives the natural conclusion from its application, namely, Dieks' modal interpretation of QM. In fact, it is the strict application of this principle within the orthodox formalism which leads Dieks to conclude that QM describes systems which have only one single real property.

\section{Dieks' Conclusion: Quantum Systems Possess Only One Real Property}

Dieks' realist modal approach to QM \cite{Dieks88a, Dieks07, Dieks10} attempts to interpret the orthodox formalism in terms of systems which possess definite valued (actual) properties. Modal interpretations are constrained by four general ideas or desiderata (see for a detailed analysis \cite{Vermaas99}). 

\begin{enumerate}
{\it \item[i.] We should stay close to the orthodox formalism of QM which has been proven to be empirically adequate. 

\item[ii.] We should attempt to provide an objective realist description of what is going on according to the theory beyond the mere reference to measurement outcomes in an instrumentalist fashion. 

\item[iii.] The proposed objective description should be in terms of systems which possess actual (definite valued) properties. 

\item[iv.] We must provide such interpretation of the orthodox formalism without ``adding anything by hand'' ---such as {\it ad hoc} rules---, respecting the symmetries and features of the mathematical structure.} 
\end{enumerate}

\noindent From these constraints Dieks derives the only possible conclusion: each quantum system possesses only one single {\it actual property}. Let us discuss in some detail his interpretation of the formalism.

According to Dieks, given a vector in Hilbert space, $\Phi $, there is only one real (actual) property related to this quantum system. In order to find this property we just need to write $\Phi$ in the specific  basis in which the state is mathematically represented as a superposition of one single term: $|\phi \rangle$. The ket $|\phi \rangle$ has a one-to-one relation to the projection operator, $|\phi \rangle \langle \phi |$, which in turn is interpreted as an {\it actual property}. Why is this property called ``actual''? Because, following Einstein's actualist definition of {\it element of physical reality},\footnote{We might recall Einstein's definition \cite[p. 777]{EPR} of an element of physical reality: ``If, without in any way disturbing a system, we can predict with certainty (i.e., with probability equal to unity) the value of a physical quantity, then there exists an element of reality corresponding to that quantity.''} this property can be predicted with certainty (probability = 1). On the contrary, the kets related to properties that appear within quantum superpositions of more than one term cannot be predicted with certainty and thus cannot be considered as being actual (or real) properties. The properties which pertain to superpositions of more than one term are called in the literature {\it indefinite properties} because ---due to certain constraints which we will analyze in the next section--- they cannot be interpreted as definite valued properties. Dieks, following a tradition that goes back to Heisenberg's interpretation and quantum logic, calls them {\it possible properties} (see for example\cite{Dieks07}).

Thus, if we ask the question: what is real according to $\Psi$? The answer provided by Dieks' modal interpretation is that there is only one {\it actual} real property, $|\phi \rangle \langle \phi |$, and many {\it possible} ones, $| K_i \rangle \langle K_i |$. According to Dieks' interpretation {\it possible properties} might also produce in the future measurement outcomes. However, strictly speaking, these properties cannot be considered as part of reality. In a more or less covered instrumentalist fashion, Dieks interprets possible properties in terms of the prediction of future measurement outcomes. But this, as mentioned above, goes against desiderata {\it (ii)} which promised an objective description of properties beyond the mere reference to measurement outcomes. As in most realist interpretations of QM, indefinite properties remain in an ontological limbo: even though they are not real, they relate to the prediction of measurement outcomes. However, this ontological distinction between a real {\it actual} property and the many {\it possible} properties, is not derived from the mathematical formalism of the theory. This distinction is imposed by a reading which ---like in the hidden variables program--- presupposes the Newtonian metaphysical stance according to which `Reality = Actuality'.\footnote{It is interesting to notice that Guido Bacciagaluppi always considered modal interpretations as a hidden variable interpretation (see \cite{Bacciagaluppi96}). Dieks was against this idea until, in a paper of 2007, changed drastically his position reconsidering modal interpretations in relation to Bohmian mechanics (see \cite{Dieks07}).}  

Dieks' modal interpretation is a coherent scheme. However, due to the (metaphysical) demand to discuss the formalism of QM in terms of `systems with definite valued properties' {\it (iii)} and the constraints imposed by Kochen-Specker (KS) type theorems to modal interpretations \cite{Bacciagaluppi95, RFD14, DFR06, Vermaas97}, Dieks is forced to conclude that quantum systems have only one real property. Dieks modal interpretation, as many proposals in the same line of quantum logic, leaves outside of the realm of physical reality the {\it indefinite} or {\it possible} properties described and probabilistically predicted by the quantum formalism. This interpretational maneuver contradicts the intuition of physicists according to which physical reality must be directly related to the predictive capacity of the formalism of the theory. As Robert Griffiths \cite[p. 361]{Griffiths02} makes the point: ``If a theory makes a certain amount of sense and gives predictions which agree reasonably well with experimental or observational results, scientists are inclined to believe that its logical and mathematical structure reflects the structure of the real world in some way, even if philosophers will remain permanently skeptical.'' 

In QM, given a vector $\Phi$, we know that all its mathematical representations, given in general by superpositions of more than one term, $\sum c_{i} | d_{i} \rangle$, which are constituted by indefinite or possible properties do provide ---even though probabilistically--- empirically meaningful predictions. So why should these properties not be considered when discussing about physical reality? It becomes evident that, leaving aside metaphysical actualism, realist interpretations should take into account not only actual properties but also possible ones when considering what is real according to the quantum formalism (see for a detailed analysis of this point \cite{deRonde16a}). As mentioned above, the reason behind this minimalist interpretation is related to the well known constraints imposed by KS type theorems to modal interpretations. We completely agree with Dieks that this is the only rigorous conclusion which can be drawn when considering, on the one hand, the orthodox formalism of QM ---through desiderata {\it (i)} and {\it (iv)}---,  and on the other, the metaphysical constraints imposed by the notions of `system' and `property' ---through desiderata {\it (ii)} and {\it (iii)}.

\section{Revisiting the Applicability of Metaphysical Identity in QM}

As obvious as it might sound, in order claim that `something is the same to itself' we need to specify what is the `something' in question which remains `the same'. In the case of classical physics, as we discussed above, before applying {\it metaphysical identity}, we need to apply PE and PNC. The first principle allows to make the claim that the `something' in question exists (PE); the second principle, that this existent is constituted by non-contradictory properties (PNC). Both of these statements face serious difficulties when related to the orthodox formalism of QM. The first difficulty relates to quantum contextuality ---understood in ontological terms through the KS theorem\footnote{See \cite{deRonde16c} for a detailed  analysis and discussion of the meaning of quantum contextuality.}--- which precludes the possibility to interpret projection operators in terms of preexistent or actual (definite valued) properties. Let us discuss this more in detail. 

In QM the frames under which a vector is represented mathematically are considered in terms of orthonormal bases. We say that a set $\{x_1,\ldots,x_n\}\subseteq {\cal H}$, an $n$-dimensional Hilbert space, is an \emph{orthonormal basis} if $\langle x_{i} | x_{j} \rangle = 0$ for $1 \leq i , j \leq n$ and $\langle x_i|x_i\rangle=1$ for all $i=1,\ldots,n$. A physical quantity is represented by a self-adjoint operator on the Hilbert space ${\cal H}$. We say that $\mathcal{A}$ is a $\emph{context}$ if $\mathcal{A}$ is a commutative subalgebra generated by a set of self-adjoint bounded operators $\{A_1,\ldots,A_s\}$ of ${\cal H}$. Quantum contextuality, which was most explicitly recognized through the KS theorem \cite{KS}, asserts that a value ascribed to a physical quantity $A$ cannot be part of a global assignment of values but must, instead, depend on some specific context from which $A$ is to be considered.

Physically, a global valuation allows us to define the preexistence of definite properties. Mathematically, a  \emph{valuation} over an algebra $\mathcal{A}$ of self-adjoint operators on a Hilbert space, is a real function satisfying,
\begin{enumerate}
\item[1.] \emph{Value-Rule (VR)}: For any $A\in\mathcal{A}$, the value $v(A)$ belongs to the spectrum of $A$, $v(A)\in\sigma(A)$.
\item[2.] \emph{Functional Composition Principle (FUNC)}: For any $A\in\mathcal{A}$ and any real-valued function $f$, $v(f(A))=f(v(A))$.
\end{enumerate}

\noindent We say that the valuation is a \emph{Global Valuation (GV)} if $\mathcal{A}$ is the set of all bounded, self-adjoint operators. In case $\mathcal{A}$ is a context, we say that the valuation is a \emph{Local Valuation (LV)}. We call the mathematical property which allows us to paste consistently together multiple contexts of {\it LVs} into a single {\it GV}, {\it Value Invariance (VI)}. First assume that a {\it GV} $v$ exists and consider a family of contexts $\{ A_i \}_I$. Define the {\it LV} $v_i:=v|_{A_i}$ over each $A_i$. Then it is easy to verify that the set $\{v_i\}_I$ satisfies the \emph{Compatibility Condition (CC)}, 
$$v_i|_{ A_{i} \cap A_j} =v_j|_{A_i\cap A_j},\quad \forall i,j\in I.$$

\noindent The {\it CC} is a necessary condition that must satisfy a family of {\it LVs} in order to determine a {\it GV}. We say that the algebra of self-adjoint operators is \emph{VI} if for every family of contexts $\{ A_i\}_I$ and {\it LVs} $v_i: A_i \rightarrow \mathbb{R}$ satisfying the \emph{CC}, there exists a {\it GV} $v$ such that $v|_{A_i}=v_i$.

If we have {\it VI}, and hence, a {\it GV} exists, this would allow us to give values to all magnitudes at the same time maintaining a {\it CC} in the sense that whenever two magnitudes share one or more projectors, the values assigned to those projectors are the same in every context. The KS theorem, in algebraic terms, rules out the existence of {\it GVs} when the dimension of the Hilbert space is greater than $2$. The following theorem is an adaptation of the KS theorem ---as stated in \cite[Theorem 3.2]{DF}--- to the case of contexts:

\begin{thm}[KS Theorem] If ${\cal H}$ is a Hilbert space of $\dim({\cal H}) > 2$, then a global valuation is not possible.
\end{thm}

\noindent KS makes reference to the interpretation of the quantum formalism in terms of systems with actual (definite valued) properties. Put in a nutshell, quantum contextuality deals with the formal conditions that any realist interpretation which respects orthodox Hilbert space QM must consider in order to consistently provide an objective physical representation of reality. These conditions preclude the possibility of interpreting any quantum state, $\Psi$, in terms of a set of definite valued properties. The principle of metaphysical identity cannot be applied to `something which possesses objective properties' in the context of QM due to the fact there is no objective way in order to define the `something' in question in terms of `possessed objective properties' (see for a more detailed analysis \cite{deRonde16c}). 

The second barrier in order to apply metaphysical identity to QM deals with the superposition principle and the existence of `Schr\"odinger cat states' of the type: $ c_{1} | \uparrow_x> + \ c_{2} | \downarrow_x>$. As we argued in \cite{daCostadeRonde13}, quantum superpositions seem to imply the need of making reference to contradictory properties. This remark precludes the very possibility to interpret superpositions in terms of metaphysical identities. Let us recall that an existent physical individual or system (PE)  must obviously posses non-contradictory properties (PNC) in order to be regarded as identical to itself (PI). According to classical metaphysics `something' cannot be in contradictory states simultaneously: `up' and `down', `dead' and `alive', etc. But according to the quantum formalism ---and due to the impossibility to interpret quantum probability in epistemic terms---, so it seems, the same `quantum particle' can posses simultaneously the property of `being decayed' and the property of `being not-decayed'; a `quantum system' can have `spin up' and `spin down' at the same time. Obviously, a system cannot be defined in terms of paraconsistent properties. One cannot apply metaphysical identity without the sound application of PNC. Contradictory properties preclude the very possibility of discussing about a physical system or individual. As we explained above, accepting contradictory properties to characterize a `quantum system' would be going against the basic metaphysical definition of what can be considered to be a `physical system'. Arguing that all this is due to the ``quantumness'' of elementary particles is just escaping the problem of properly accounting and making sense of such quantum realm. Saying that things are weird because they are quantum is not providing a proper physical representation. 

As we discussed in the previous section, if one applies the modal interpretation's constraints, {\it (i)} and {\it (iv)}, to stay close to the orthodox  formalism of QM, and to describe physical reality {\it (ii)} in terms of systems with definite valued properties {\it (iii)} ---implicitly applying PE, PNC and PI---; then one cannot escape the natural conclusion derived by Dieks: {\it A vector in Hilbert space represents reality objectively in terms of one single property.} However, there are difficulties with such an interpretation of QM. 

Firstly, a system which possess only one property seems to be not a very interesting physical system. It is even difficult to imagine what a single property system would be like. Furthermore, the reference to possible properties remains completely unclear. 

Secondly, this interpretation obscures the richness of the formalism. On the one hand, it bypasses the specificity of the multiple mathematical representations of an abstract vector as well as their contextual character ---discussed in the previous sections. On the other hand, the distinction between different levels of mathematical representation is overlooked due to an interpretation which makes an implicit application of metaphysical identity. 

Thirdly, even though the so called indefinite or possible properties are the main elements of new quantum technologies ---due to metaphysical actualism--- their relation to physical reality remains  completely unspecified. Such a description seems completely incapable to explain ---independently of measurement outcomes--- the existence of quantum superpositions and entanglement, two of the most important features within quantum information processing. Quantum superpositions are being tested today in the lab. Despite the many interpretations which attempt to deny their physical existence, Schr\"odinger cats are getting fat  and bigger \cite{Blatter00, Nature15, NimmrichterHornberger13}.\footnote{In this respect, it should become clear that a necessary requirement for discussing the development of quantum information processing is to posses a rigorous definition of quantum superpositions such as the one proposed in \cite{deRonde16b}.}

\section{Final Remarks}

Michel Bitbol \cite[p. 72]{Bitbol10} makes the following interesting remark: ``The tendency to reify state vectors [in QM] manifests itself in the use of the very word `state'. The `grammar' (in Wittgenstein's sense) of the word `state' requires that this is the state of something; that it belongs to something; that it characterizes this something independently of anything else. Such grammar, and the conception associated to it, is sufficient to generate one of the major aspects of the measurement problem.'' In this paper, from a realist perspective, we are certainly not criticizing the reification of vectors. A realist obviously needs to explain in which sense the formalism relates to a particular representation of reality. What we have attempted to show is that the grammar imposed by the word `state' and `system' is simply not adequate to interpret QM. This does not invalidate realism itself ---as many, including Bitbol seem to argue---, it simply makes explicit the inadequacy of the classical metaphysical scheme when related to the orthodox formalism of QM. This does not preclude the possibility to develop new metaphysical schemes that would allow us to explain what is QM really talking about. 

In classical physics the mathematical structures that are used are based on classical logic. But classical logic is also one of the formal cornerstones of the metaphysics of entities. We know that the mathematical structure of QM has features which go clearly against the basic principles of classical logic, and in particular, the principle of identity. Thus, it would be then a strong surprise, that such non-classical mathematical syntactic structure could be interpreted following the same semantic and classical concepts that have been used to interpret classical physics. It seems to us clear that, from a philosophical perspective, we should not expect {\it a priori} this to be necessarily the case. 

We are convinced that imposing uncritically a metaphysical interpretation that is simply not coherent with the formalism will not help us in understanding what the theory is really talking about. On the very contrary, as we have attempted to show in this paper, assuming improper metaphysical schemes in the semantic level will certainly obscure the features already present within the formalism itself in the syntactic level. It is in fact these non-classical features which, we believe, are the main elements which should be considered when attempting to develop a coherent interpretation of QM that respects the orthodox formalism. In short, we should learn from the mathematical formalism of the theory instead of uncritically imposing, implicitly (as in the orthodox interpretation) or explicitly (as in the hidden variable project), a classical metaphysical scheme based on {\it metaphysical identity}.  

The relation between a mathematical formalism (in the syntactic level) and a set of physical concepts (in the semantic level) cannot be considered ---in general--- as ``self evident''. Physical concepts can not be ``read out literally'' from the formalism ---as some authors might claim. Physics is neither committed to classical concepts and language ---as others believe.\footnote{Our statement goes clearly against the Bohrian \cite[p. 7]{WZ} stance according to which: ``[...] the unambiguous interpretation  of any measurement must be essentially framed in terms of classical physical theories, and we may say that in this sense the language of Newton and Maxwell will remain the language of physicists for all time.'' In this respect, also according to Bohr [{\it Op. cit.}], ``it would be a misconception to believe that the difficulties of the atomic theory may be evaded by eventually replacing the concepts of classical physics by new conceptual forms.''} Quite on the contrary, the history of physics is also the history of the creation of new physical concepts. The authors of this paper believe that the concept of `physical object' or `physical system' is a notion that was created and developed through the history of physics; it is not a {\it true concept} that we have discovered in a Platonic heaven; and neither it is a concept that must necessarily constrain all our present and future physical theories. In particular, as we have argued, it might not constrain QM.

\section*{Acknowledgements}

This paper was presented in the {\it 18th UK European Conference on Foundations of Physics}, July 2016. We would like to thank the audience for useful discussion. This work was partially supported by the following grants: FWO-research community W0.030.06. CONICET RES. 4541-12 and the Project PIO-CONICET-UNAJ (15520150100008CO) ``Quantum Superpositions in Quantum Information Processing''.

\end{document}